\documentclass[pdflatex,sn-nature]{sn-jnl}% Style for submissions to Nature Portfolio journals
\usepackage{graphicx}%
\usepackage{multirow}%
\usepackage{amsmath,amssymb,amsfonts}%
\usepackage{amsthm}%
\usepackage{mathrsfs}%
\usepackage[title]{appendix}%
\usepackage{xcolor}%
\usepackage{textcomp}%
\usepackage{manyfoot}%
\usepackage{booktabs}%
\usepackage{algorithm}%
\usepackage{algorithmicx}%
\usepackage{algpseudocode}%
\usepackage{listings}%
\theoremstyle{thmstyleone}%
\theoremstyle{thmstyletwo}%

\theoremstyle{thmstylethree}%

\begin{document}

\title[Article Title]{Coherent Control of High-Order Lightwave Currents in Gate-Tunable Bilayer Graphene} 
%Gate-Tunable Emergence and Coherent Control of Exceptional Multi-Order Terahertz Photocurrent in Bilayer Graphene
%Femtosecond Coherent Control and Multi-Order Correlation Tomography of Gate-Tunable Lightwave Photocurrents in Bilayer Graphene

%Nonequilibrium strong-field interometric THz lightwave photocurrent 
%{Femtosecond–Terahertz Photocurrent Control in Graphene via Gate-Tunable Lightwave Acceleration of Dirac Electrons}
%nonlinear interfrometric THz photocurrent spectroscopy 

\author[1,2]{\fnm{Samuel} \sur{Haeuser}}%\email{iauthor@gmail.com}

\author[2]{\fnm{Martin} \sur{Mootz}}%\email{iiauthor@gmail.com}

\author[3]{\fnm{Dasol} \sur{Kim}}%\email{iiiauthor@gmail.com}

\author[1,2]{\fnm{Avinash} \sur{Khatri}}%\email{iiiauthor@gmail.com}

\author[2]{\fnm{Joong-Mok} \sur{Park}}%\email{iiauthor@gmail.com}

%\author[3]{\fnm{Junseok} \sur{Seo}}%\email{iiiauthor@gmail.com}

\author[1,2]{\fnm{Randall K.} \sur{Chan}}%\email{iiauthor@gmail.com}

\author[2]{\fnm{Richard H. J.} \sur{Kim}}%\email{iiauthor@gmail.com}

\author[2]{\fnm{Yongxing} \sur{Yao}}%\email{iiauthor@gmail.com}

\author[4]{\fnm{Ilias E.} \sur{Perakis}}%\email{iiiauthor@gmail.com}

\author[5]{\fnm{Hou-Tong} \sur{Chen}}%\email{iiiauthor@gmail.com}

\author[1,2]{\fnm{Thomas} \sur{Koschny}}%\email{iiiauthor@gmail.com}

\author[3]{\fnm{Long} \sur{Ju}}%\email{iiiauthor@gmail.com}

\author[1,2*]{\fnm{Jigang} \sur{Wang}} %\email{jgwang@ameslab.gov}

\affil[1]{\orgdiv{Department of Physics and Astronomy}, \orgname{Iowa State University}, \orgaddress{\street{Street}, \city{Ames}, \postcode{50011}, \state{Iowa}, \country{USA}}}

\affil[2]{\orgdiv{Ames National Laboratory}, \orgname{U.S. Department of Energy}, \orgaddress{\street{Street}, \city{Ames}, \postcode{50011}, \state{Iowa}, \country{USA}}}

\affil[3]{\orgdiv{Department of Physics}, \orgname{Massachusetts Institute of Technology}, \city{Cambridge}, \state{Massachusetts}, \country{USA}}

\affil[4]{\orgdiv{Department of Physics}, \orgname{University of Alabama at Birmingham}, \city{Birmingham}, \state{Alabama}, \country{USA}}

\affil[5]{\orgdiv{Center for Integrated Nanotechnologies}, \orgname{Los Alamos National Laboratory}, \city{Los Alamos}, \state{New Mexico}, \country{USA}}

\affil[*]{Address correspondence to: jgwang@ameslab.gov; jgwang@iastate.edu}


\abstract{The sub-cycle acceleration of charge carriers in an electronic circuit by tailored terahertz (THz) electric-field waveforms, rather than conventional pulsed bias voltages, opens a transformative frontier for lightwave electronics operating far beyond the scaling limits of gigahertz electronic technologies. Yet gate-tunable coherent control of multi-order and non-perturbative currents has remained elusive. Here, we demonstrate lightwave control over the amplitude, phase flips, and zero-crossings of THz field-driven high-order currents by utilizing phase-locked pairs of single-cycle THz fields in a dual-gated bilayer graphene (BLG) device. By leveraging the in-situ tuning of both bandgap and charge doping, we reveal extreme THz photocurrent rectification and sub-THz-cycle current generation that are governed by a strong field-driven crossover from interband multi-photon coherence to intraband lightwave acceleration, whose competition provides an exceptional sensitivity to the THz-pulse pair drive. Our interferometric nonlinear THz photocurrent spectroscopy of sub-cycle electronic excitations accesses and controls non-perturbative, multi-quantum photocurrents, which manifest as an extreme-order THz frequency comb. 
Quantum-kinetic simulations map these lightwave-controlled electron trajectories, revealing that sub-THz-cycle ballistic transport underpins gate- and THz-field-driven functionality enabled by electrostatic doping and bandgap tuning within the atomically thin device architecture. The demonstrated on-demand coherent control of nonperturbative photocurrent transport regimes can be extended to a broad range of atomically thin device architectures, opening new avenues for gate- and field-programmable logic, memory, and quantum simulators in which electronic pathways are steered on sub-cycle, petahertz timescales and atomic length scales.

}

\maketitle

\section{Introduction}

%The Rise of Lightwave Electronics and the Photocurrent Paradox
%The ability to steer electrons using the oscillating electric field of light, rather than traditional bias voltages, has established a new control principle for lightwave current engineering~\cite{Borsch2023,Heide2024NatRevPhys},
%birthed the field of ``lightwave electronics,” 
%promising a leap from gigahertz to THz communication and information processing. 
By driving electrons ballistically within a single THz light cycle, remarkably nonlinear and non-perturbative phenomena, assisted by lightwave currents, emerge across diverse material classes: from high-harmonic generation in semiconductors and atomically thin crystals~\cite{Ghimire2011NatPhys,Schubert2014NatPhoton,Hohenleutner2015Nature,Langer2017NatPhoton,Liu2017NatPhys,GhimireReis2019NatPhys}, to sub-cycle motion control of massless carriers in topological insulators~\cite{Reimann2018Nature,Schmid2021Nature}, and even the manipulation of Higgs modes~\cite{Luo2023NatPhys,huang2025discovery} and pseudospin precession~\cite{yang2019lightwave,vaswani2019discovery} in superconductors~\cite{Huang2026}. 
%In semiconductors and atomically thin crystals, this concept has enabled non-perturbative high-harmonic generation and direct tracking of ultrafast carrier trajectories, including landmark demonstrations in bulk crystals and monolayer semiconductors. In Dirac materials such as graphene, strong-field driving has further revealed highly tunable quantum pathways and harmonic responses linked to the motion of massless carriers. In superconductors, lightwave acceleration extends this paradigm to paired condensates, producing pseudo-spin precession and high-harmonic responses forbidden by equilibrium symmetry. Together, these studies show that lightwave fields can reshape centre-of-mass momentum, transient symmetry breaking and topological selection rules on ultrafast timescales.
Such a lightwave current engineering promises a leap from gigahertz to THz communication and information processing.
%In these systems, the lightwave acts as a non-contact ``electrode,” controlling the center-of-mass momentum and symmetry-breaking in ways that are fundamental ly linked to the material’s topology. 
However, a critical gap remains: while optical emissions routinely access extreme-order nonlinear regimes, photocurrents--the most direct transduction of light into electrical signals--almost exclusively access conventional, second-order perturbative processes and incoherent relaxation~\cite{Ma2023,Xu2018}. 
The realization of lightwave photocurrents with extreme nonlinearity has been hindered by a formidable multi-physics challenge: high-order THz current generation and detection involves a complex interplay of sub-cycle field acceleration, interband coherence, intraband ballistic transport, and fast charge collection before incoherent scattering and relaxation take hold.
%Observing these high-order responses is therefore not merely a quest for ``larger” signals, but a necessary probe into the multi-physics of strong-field dynamics in microelectronics.

%The Bottleneck: Static Observations vs. Active Gate Control
Another outstanding scientific challenge for lightwave transport technology is the gate tunability of high order, THz field-driven currents. Prior nonlinear photocurrent observations are largely passive, determined by the intrinsic properties of the chosen nonlinear crystal.
%with some control via the waveform of the driving THz pulses. 
To move toward gate-programmable lightwave functionality, this requires a photocurrent platform where the extreme THz nonlinearity itself (order, amplitude, and direction% maybe "phase" is better than "direction"
) is electrically gate-tunable by the in situ control of bandgap, carrier density, and symmetry breaking properties. Gate-controlled photocurrent studies have commonly been done at visible\cite{vogt2020ultrafast,sun2012ultrafast} or infrared frequencies~\cite{Ma2023,Xu2018,Ju2017Science,Xu2010NanoLett,Gabor2011Science,han2021accurate}. Recently THz photocurrents have accessed perturbative and photo-thermoelectric processes~\cite{krishna2025terahertz, xi2025terahertz,zhou2026gate}, rather than high-order, non-perturbative THz currents induced by coherent-lightwave-driven processes. 
This is partly due to the experimental difficulty of detecting THz photocurrent in gated atomically thin devices and isolating multi-quantum coherent photoresponses from the dominant photothermal background.  
Although coherent nonlinear THz spectroscopy using phase-locked pulses has emerged as a powerful tool to resolve lightwave acceleration~\cite{Mootz2022CommunPhys,Luo2023NatPhys,Huang2026}, its application to gated nanodevice architectures remains unexplored. Similarly, combining conventional Fourier-transform photocurrent spectroscopy with ultrafast lasers has primarily probed incoherent photocurrent relaxation and has not accessed non-perturbative and high-order current generation. 
We bridge this gap by using phase-locked, single-cycle THz pulses to coherently drive extreme nonlinear photocurrents and simultaneously measure sub-cycle-resolved nonlinear current interferometry, tracing the non-perturbative THz currents under active electrostatic gate control.
%measure the spectra and electric pathways of THz photocurrents 

%Bilayer Graphene: The Ideal Quantum Laboratory
%Atomically thin van der Waals (vdW) materials such as 
Bilayer graphene (BLG) provides a compelling platform for implementing gate-tunable, coherent-lightwave-driven currents. While monolayer graphene exhibits interesting high-harmonic optical emission~\cite{Yoshikawa2017Science,Hafez2018Nature} and THz electrodynamics~\cite{Lui2011,Li2012} due to its conical band structure, the absence of a bandgap even under electrostatic gating limits the control due in part to unwanted dark currents. In contrast, BLG allows for the independent electrostatic tuning of both carrier density via the Fermi energy ($E_\mathrm{F}$) and displacement-field-induced bandgap ($\Delta$)~\cite{Zhang2009Nature,Taychatanapat2010PRL,Ohta2006,Zhou2022, DeLaBarrera2022,Cao2018a}. 
Dual-gate control, combined with a Dirac-derived band structure away from the zone center and ultrafast optical response through weak screening, creates a unique setting in which strong THz electric fields can drive carriers into non-perturbative regimes controlled by $E_\mathrm{F}$ and  $\Delta$. 
Although BLG has already enabled gate-tunable valley and excitonic phenomena at infrared frequencies~\cite{Ju2017Science,Yin2022Science}, these demonstrations have remained confined to perturbative responses, and THz lightwave currents have never been observed.  
%and become strongly intertwined.   

We use the phase-locked pair of intense single-cycle THz pulses, in combination with a dual-gated BLG device, to generate and resolve the gate-tunable emergence of
%of high-order lightwave currents . 
%The phase-locked pairs of intense single-cycle THz driving in gated-BLG structures 
%This control scheme allows us to transcend prior boundaries and resolve 
extreme, non-perturbative lightwave photocurrents. The strong THz electric field drives carriers substantially away from the band edge into a highly nonlinear conical band where interband transitions and intraband ballistic acceleration strongly compete, leading to sensitive field-dependent control. By leveraging electrostatic gating and sub-cycle interferometric excitation and detection of THz pulse-pairs, we demonstrate that the extreme-order THz currents can be driven deterministically and controlled at terahertz frequencies. These findings successfully bridge the divide between strong-field quantum dynamics and atomically thin device functionalities, laying the groundwork for gate-tunable lightwave electronics.

\begin{figure}[!ht] 
    \centering
    \includegraphics[width=0.9\linewidth]{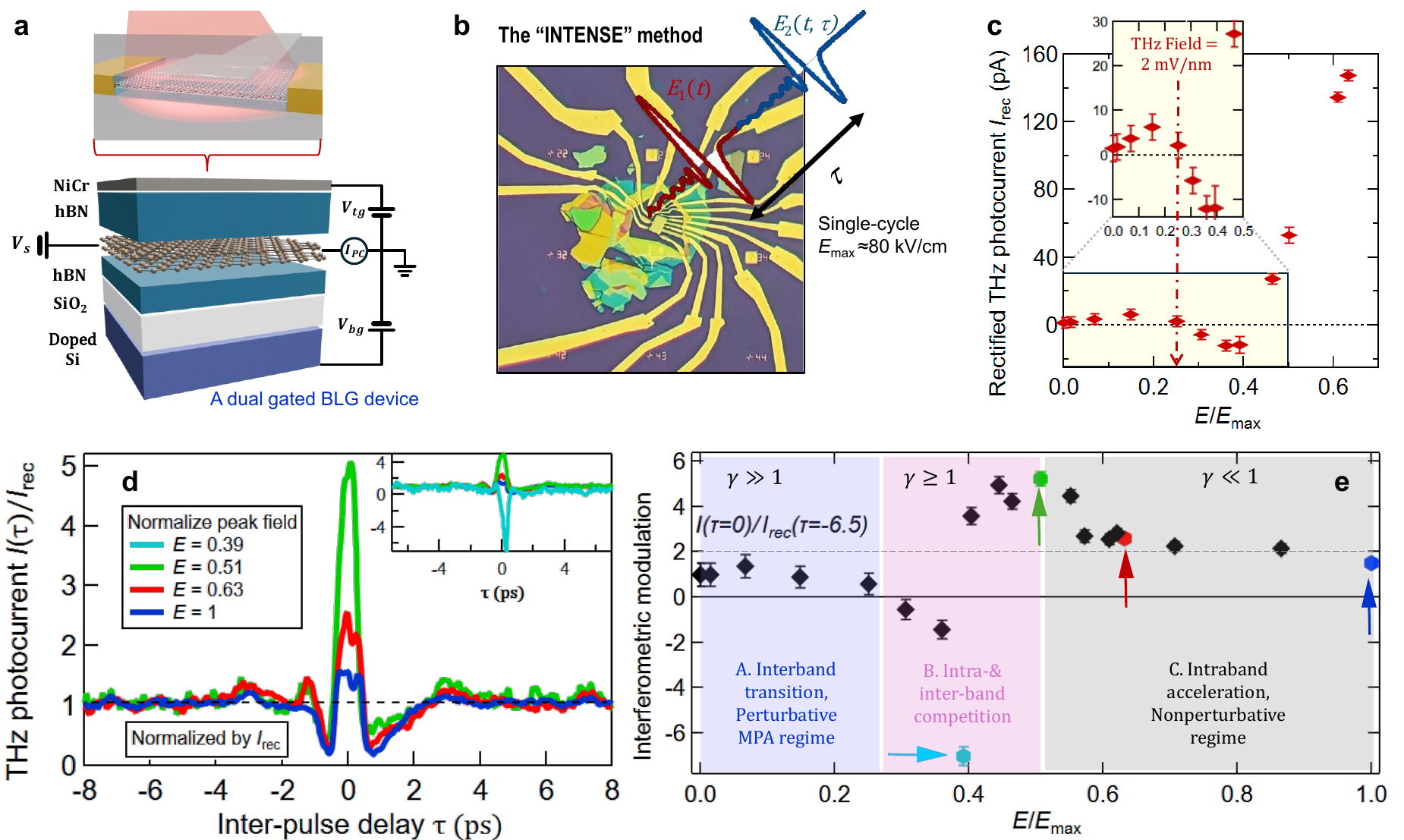}
    \caption{\textbf{Sub-cycle terahertz lightwave transport and non-perturbative photocurrent rectification in bilayer graphene.} 
    \textbf{a}, Cross-section of the dual-gated bilayer graphene (BLG) heterostructure, detailing the independent top gate ($V_{\mathrm{tg}}$), global doped-Si back gate ($V_{\mathrm{bg}}$), hBN encapsulation layers, and metal electrodes.
    \textbf{b}, Architecture of the ``INTENSE'' (interferometric terahertz nonlinear spectroscopy) method, where phase-locked single-cycle pulses, $E_1(t)$ and $E_2(t,\tau)$, are focused onto the BLG device as a function of interpulse delay, $\tau$.
    \textbf{c}, Steady-state rectified THz photocurrent, $I_{\mathrm{rec}}$, as a function of the normalized THz field amplitude $E/E_{\mathrm{max}}$ ($E_{\mathrm{max}} \sim 80~\mathrm{kV\,cm^{-1}}$); the inset highlights the low-field rectification.
    \textbf{d}, Normalized time-domain interferograms, $I_{\mathrm{THz}}(\tau)/I_{\mathrm{rec}}$, at representative peak fields: $E/E_{\mathrm{max}} = 0.39$ (cyan, inset), $0.51$ (green), $0.63$ (red), and $1.0$ (blue). The inset isolates the sub-cycle phase reversal that emerges in the intermediate field regime.
    \textbf{e}, The extracted interferometric modulation ratio, $I(\tau=0)/I_{\mathrm{rec}}(\tau=-6.5~\mathrm{ps})$, mapped across the electric field continuum. Shaded backgrounds delineate three distinct Keldysh transport regimes: the perturbative multiphoton absorption (MPA) regime ($\gamma \gg 1$, blue background), the interband--intraband competition ($\gamma \geq 1$, pink background) containing the negative-ratio phase flip (cyan marker), and the non-perturbative lightwave acceleration regime ($\gamma \ll 1$, gray background). Colored arrows indicate the field positions corresponding to the time-domain traces in (\textbf{d}).}
    \label{fig:figure1}
\end{figure}

\section{Strong-field THz photocurrent rectification in bilayer graphene}

%We study the measured photo-conductive response in fabricated bi-layer graphene (BLG) devices by driving the system with two intense near-single cycle THz pulses. The 2D inter-pulse delay is precisely controlled to drive the system as a function of interference at the THz frequencies. Figure 1 (c) depicts the intense THz pulses with an interpulse delay, $\tau$, on the BLG device. 

%we term INTENSE (Interferometric Nonlinear THz Excitation Spectroscopy of Extreme photocurrents). 
As illustrated in Fig.~\ref{fig:figure1}a, the electrostatic profile of the encapsulated BLG channel in our device is governed by two independent voltage inputs: the top gate ($V_{\mathrm{tg}}$) and the back gate ($V_{\mathrm{bg}}$).  
%These gates apply independent, perpendicular electric fields from above and below the van der Waals heterostructure, respectively.
By scaling these inputs into individual gate displacement fields ($\delta D_{\mathrm{tg}}$ and $\delta D_{\mathrm{bg}}$), we can independently tune the total carrier density ($n$, representing doping) and the net perpendicular displacement field ($D$, representing the band gap-opening field) (Supplementary Information, Section 1).  
%through the following linear combinations:$$n = \frac{1}{e}(D_{\mathrm{tg}} + D_{\mathrm{bg}})$$$$D = \frac{1}{2}(D_{\mathrm{tg}} - D_{\mathrm{bg}})$$where $e$ is the elementary charge. 
We characterize the high-order nonlinear photoconductive response of our BLG devices using INTENSE (Interferometric Nonlinear THz Excitation of Non-perturbative Sub-cycle Electronics), a method that highlights its capability of resolving non-perturbative lightwave currents driven by intense THz pulse-pair fields. As illustrated in Fig.~\ref{fig:figure1}b, the dual-gated BLG device is exposed to a pair of phase-locked, single-cycle THz pulses with comparable peak fields of up to $E_{\mathrm{max}}\approx80~\mathrm{kV~cm}^{-1}$, where adjusting the interpulse delay, $\tau$, yields sub-cycle femtosecond phase control in electric transport measurements (Supplementary Information, Section 2). 
The INTENSE method extends conventional Fourier-transform photocurrent spectroscopy~\cite{Ross2014, Ju2014, Ju2017Science} from white-light and infrared interferometry to the THz strong-field non-perturbative regime. 
Whereas conventional photocurrent interferometry has only observed linear absorption and incoherent ultrafast relaxation, the strong field-driven THz regime collapses the perturbative hierarchy and thereby enables the observation and coherent control of extreme high-order photocurrent contributions on sub-cycle timescales.
%far exceeding standard linear-in-intensity responses. 
%where photocurrents enter a strictly non-perturbative state on sub-cycle femtosecond timescales.

We first present the integrated rectified photocurrent to demonstrate the strong-field-driven transport regime. We plot $I_{\mathrm{rec}}$ driven by a single THz pulse as a function of the normalized electric field amplitude ($E/E_{\mathrm{max}}$) in Fig.~\ref{fig:figure1}c. The driving pulse is centered at $\omega_{0}\sim 0.6~\mathrm{THz}$ ($\sim 2.5~\mathrm{meV}$). Through gate tuning, a displacement-field-induced bandgap of $\Delta \approx 3.5~\mathrm{meV}$ is opened--slightly exceeding $\omega_{0}$
%the photon energy of the fundamental drive
--alongside a slight electron doping of the BLG channel. 
Strong field-driven current emerges as the $E$-field amplitude increases distinct from standard linear-in-intensity photocurrent responses. Initially, the rectified photoresponse grows monotonically up to $E/E_{\mathrm{max}} \sim 0.25$, at which point the current exhibits a striking phase reversal with respect to the driving THz direction at a threshold field of $\approx 2~\mathrm{mV\,nm^{-1}}$ (shaded region, Fig.~\ref{fig:figure1}c). As the field strength increases further, the direction of the rectified transport aligns back with the driving field, accompanied by a rapid, non-monotonic increase of the total integrated photocurrent. 
We emphasize two defining attributes that underscore the unique nature of this THz-field-driven rectification current. First, the field-induced phase flip constitutes a definitive hallmark of entering the non-perturbative, strong-field-driven regime; this behavior is phenomenologically consistent with d.c. current reversals previously observed in monolayer graphene driven by phase-locked near-infrared two-cycle pulses ($\hbar\omega \approx 1.5~\mathrm{eV}$)~\cite{Higuchi2017Nature}. 
Second, remarkably, the threshold for this coherent current reversal occurs at a peak THz field strength of $2~\mathrm{mV\,nm^{-1}}$--a thousand-fold reduction compared to the $2~\mathrm{V\,nm^{-1}}$ excitation fields required in the optical domain for monolayer graphene~\cite{Higuchi2017Nature}.

%by sub-optical-cycle Landau–Zener–Stückelberg interference repeated Landau–Zener transitions on the femtosecond timescale. 

Moving beyond the single-pulse rectification that maps the strong-field THz transport regime (Fig.~\ref{fig:figure1}c), we deploy the INTENSE methodology to uncover underlying sub-cycle photoconductive dynamics that are otherwise inaccessible to time-integrated measurements. Figure~\ref{fig:figure1}d displays the cross-correlation THz photocurrent interferograms, $I(\tau)$, measured across the BLG device using THz pulse-pair excitation at various field strengths. Strikingly, the temporal profile of the interferogram varies dramatically with the driving field amplitude, as demonstrated by the three traces at $E/E_{\mathrm{max}} \approx 1$ (blue), $0.63$ (red), and $0.51$ (green), each normalized to its respective rectification current ($I_{\mathrm{rec}}$) . 
Here, the baseline current $I_{\mathrm{rec}}$ is captured at an interpulse delay of $\tau = -6.5~\mathrm{ps}$, where the two single-cycle THz pulses are completely separated in time. Furthermore, at a lower field amplitude of $E/E_{\mathrm{max}} \approx 0.39$ (cyan), the measured interferometric photocurrent undergoes a distinct phase flip relative to the driving pulse phase (Fig.~1d, inset), in clear contrast to conventional perturbative photocurrent signatures. 

For a standard linear-in-intensity photoresponse, the interferometric rectification current $I(\tau)$ is strictly proportional to the second-order correlation function, described by $S_1(\tau)=\int_{-\infty}^{\infty}(E_1(t)+E_2(t-\tau))^2 \, dt$, where $E_{1,2}(t)$ represent two similar time-dependent driving fields separated by $\tau$ (Fig.~\ref{fig:figure1}b). 
In this conventional interferometric photocurrent regime, the normalized profile of $I(\tau)$ remains invariant regardless of the driving pulse's temporal waveform or field strength, because $S_1(\tau)$ is uniquely dictated by the power spectrum of the excitation field and carries no information regarding its underlying temporal profiles. 
%and carries no information regarding its underlying temporal waveform. 
We confirm this invariance in a control experiment at a driving mid-infrared frequency of $75~\mathrm{THz}$, where nearly identical $I(\tau)$ profiles are observed across a wide range of peak fields up to $\sim 500~\mathrm{mV\,nm^{-1}}$ (Supplementary Information, Section 3). Consequently, the field-dependent $I(\tau)$ evolution and phase flips in Fig.~\ref{fig:figure1}d provide definitive evidence of quantum interference between high-order lightwave currents driven into the non-perturbative regime. This unveils hidden temporal dynamics resolved by the INTENSE methodology, yielding information entirely distinct from conventional photocurrent interferograms~\cite{Ross2014, Ju2014, Ju2017Science}.

%where $E_1$ and $E_2$ are the real time, t, dependent similar driving fields separated by an inter-pulse delay, $\tau$. 
%In most cases, this scenario holds true, allowing us to disregard all higher order correlation functions of a given system. 
%observed in our THz photocurrent interferograms 

To track these field-dependent temporal profiles more quantitatively, we define the interferometric modulation ratio, $I(\tau=0)/I_{\text{rec}}$, which evaluates the photo-response at zero time delay ($\tau = 0~\text{ps}$) against the baseline response of completely separated pulses ($\tau = -6.5~\text{ps}$, Fig. ~\ref{fig:figure1}d). 
As shown in Fig.~\ref{fig:figure1}e, this metric directly exposes three distinguishing features of strong-field electronic transport.
First, the modulation ratio varies strongly as a function of the normalized terahertz field, $E/E_{\text{max}}$, reaching a magnitude more than six times that of the background rectification current. In the perturbative, linear-in-intensity regime, this modulation ratio is strictly constrained to $S_1(\tau=0)/S_1(\tau \ll 0) \approx 2$ for identical pulse pairs, a baseline ratio explicitly confirmed by our mid-infrared control experiments on the same BLG device (Supplementary Information). The profound deviation of our THz modulation ratio from this second-order correlation limit (gray dashed line, Fig.~\ref{fig:figure1}e) indicates the emergence of high-order, coherent-lightwave-driven photocurrents, corroborating our conclusion of non-perturbative strong-field THz transport.
%that cannot be described by the standard first and second correlation functions. 
%and if $E=E_1=E_2$, at $\tau = 0$, $S = 4 \int^{\infty}_{\infty} E^2$ where as when $\tau$ is far from overlap $S = 2 \int^{\infty}_{\infty} E^2$, or in other words, 
Second, the modulation ratio exhibits clear phase flips and zero-crossings at distinct field regimes. At low driving fields, the in-phase current response grows gradually within a first region (blue), culminating in a phase flip at a threshold of $E/E_{\text{max}} \sim 0.25$. As the field increases to intermediate amplitudes, a sharp transition occurs between out-of-phase and in-phase driving, mapping out a second region (pink) characterized by high peak ratios and an additional zero-crossing point. Finally, at high fields exceeding $E/E_{\text{max}} \sim 0.52$--which defines a third region (gray)--the peak ratio declines and asymptotically approaches a value of $\approx 2$. However, this high-field return to the baseline ratio occurs via a highly unconventional interferogram line shape (plateau-like interferometric peak in the blue trace, Fig.~\ref{fig:figure1}d), pointing to a complex quantum interference involved. 
%indicating high order non-nonlinearities, 
Third, the drastic changes of the modulation ratio with the driving field, shown in Fig.~\ref{fig:figure1}e, indicate the tunability of the three photocurrent transport regimes. The underlying quantum kinematics is classified using the dimensionless Keldysh parameter, $\gamma$, adapted for a gapped two-band system~\cite{Heide2021}, $\gamma = {\Delta}/{2\hbar\Omega_{\mathrm{R}}}$ where $\Delta \propto D$ is the displacement-field-induced bandgap and $\Omega_{\mathrm{R}}\propto E_0$ is the driving-field-dependent Rabi frequency (Supplementary Information). 
%$\Omega_R = v_F e E_0 / (\hbar \omega)$ being the Rabi frequency and $v_F$ the Fermi velocity
Intriguingly, the field-tunable nonlinear photocurrent in BLG demonstrated in Fig.~\ref{fig:figure1}e undergoes a distinct THz-field-driven crossover from an interband-dominated multiphoton regime ($\gamma \gg 1$, blue) to an intraband-dominated lightwave-acceleration regime  ($\gamma \ll 1$, gray). These dramatic modulation changes manifest as phase flips and zero-crossings, serving both as a direct experimental signature of the competition between these two photocurrent channels ($\gamma \geq 1$) and as a sensitive control knob for high-order nonlinear currents.
%Multi-photon regime ($\gamma \gg 1$): Dominated by conventional multi-photon absorption and discrete interband transitions across the electrostatic bandgap $\Delta$.Crossover regime ($\gamma > 1$): An intermediate state where interband multi-photon pathways and intraband ballistic acceleration are comparable and strongly compete.Lightwave acceleration regime ($\gamma \ll 1$): The non-perturbative strong-field domain dominated by sub-cycle electron intraband acceleration, giving rise to extreme high-order coherence.
%The crossing point of this phase corresponds to the onset of intra-band electron acceleration, competing with lower transitions from inter-band multi-photon absorption. When the dominating transition flips between these two competing nonlinearities.
%Notably, due to the change in field strength through the interferogram measurement process, examining the change from the peak overlap position to the non interacting pulse separated position provides insight to the underlying electronic phenomena. 
%To better highlight the different excitation processes, Fig.~1E maps out this response through measuring the peak overlap to a non-interacting separated location ($\tau$ = -6.5 ps position) as a function of THz intensity, normalizing to the signal from the driving field $E/E_{\mathrm{max}}$ . 

\section{Strong field-driven spectral crossover from interband to intraband transport}

\begin{figure}[!ht] 
    \centering
    \includegraphics[width=0.95\linewidth]{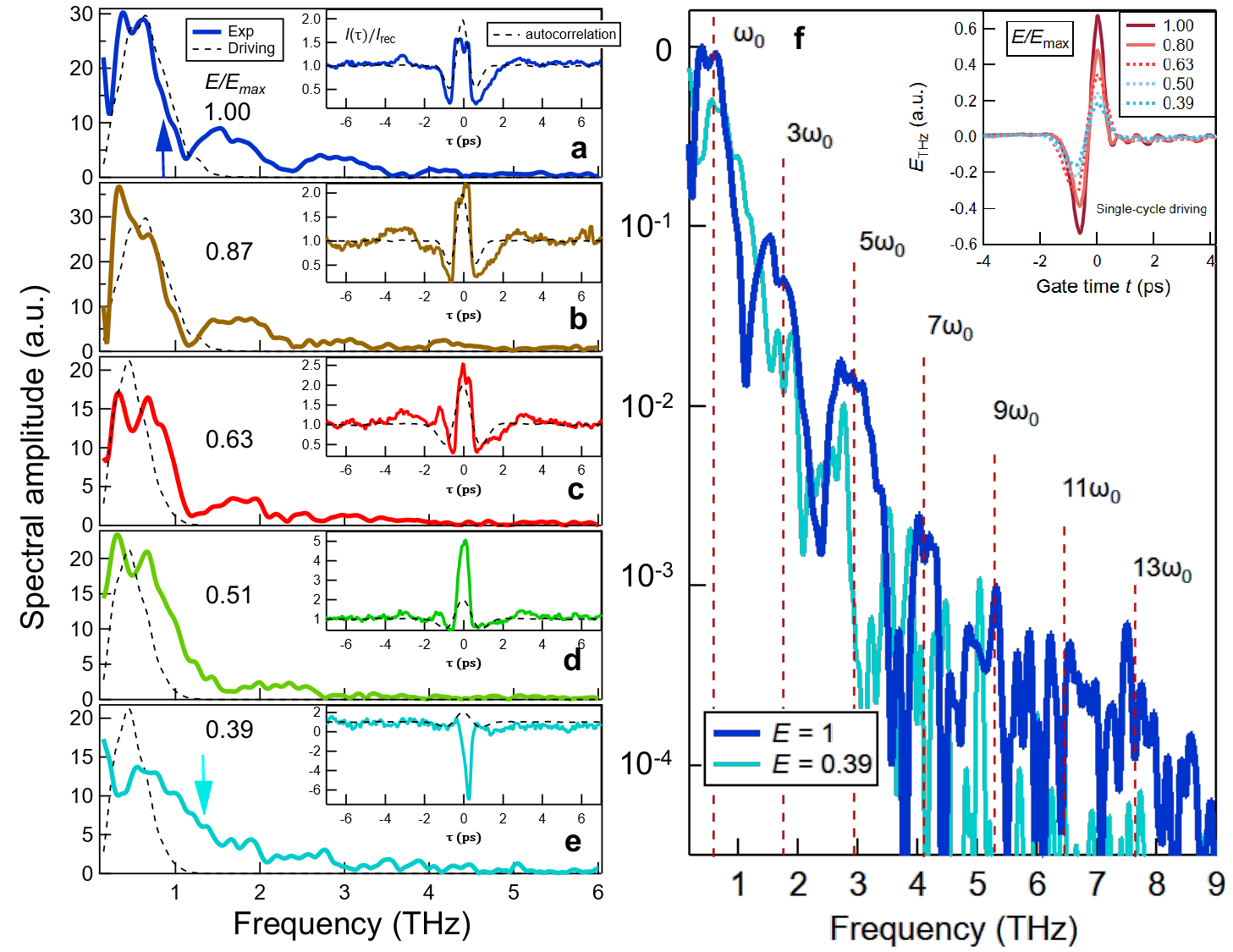}
    \caption{\textbf{Field-induced evolution and high-order multi-quantum coherence spectra in bilayer graphene.} 
    \textbf{a--e}, Photocurrent spectral amplitudes (solid colored lines) obtained via Fourier transformation of the time-domain cross-correlation interferograms (insets, solid lines) at representative normalized driving fields: $E/E_\mathrm{max} = 1.0$ (\textbf{a}), $0.87$ (\textbf{b}), $0.63$ (\textbf{c}), $0.51$ (\textbf{d}), and $0.39$ (\textbf{e}). The calculated second-order field correlation spectrum of the driving pulse is plotted for reference in each panel (black dashed lines). Insets display the corresponding real-time delay traces alongside their standard linear autocorrelation profiles (black dashed lines).
    \textbf{f}, High-sensitivity log-scale comparison of the photocurrent spectra between the low-field crossover state ($E/E_\mathrm{max} = 0.385$, cyan) and the intense strong-field limit ($E/E_\mathrm{max} = 1.0$, blue). Vertical red dashed lines indicate the precise positions of the odd-order harmonics ($\omega_0$ up to $13\omega_0$) of the fundamental drive frequency. Inset: Temporal profiles of the single driving terahertz pulses, $E_{\mathrm{THz}}(t)$, recorded at various field strengths via external electro-optic sampling (EOS).}
    \label{fig:figure2}
\end{figure}

Figure~\ref{fig:figure2} details the spectral evolution of the anomalous photocurrent interferograms across the strong-field transport regimes identified in Fig.~\ref{fig:figure1}. Fourier transformation of the time-domain cross-correlations, $I(\tau)$ (solid lines, Fig.~\ref{fig:figure2}, a to e, insets), reveals a clear restructuring of the photocurrent spectrum with increasing THz driving field (solid lines, Fig.~\ref{fig:figure2}, a to e). The corresponding single-cycle THz driving waveforms, measured externally in free space by electro-optic sampling, are shown in the inset of Fig.~\ref{fig:figure2}f. For comparison, the calculated second-order optical correlation function $S_1(\tau)$ is plotted for each field amplitude (dashed lines, Fig.~\ref{fig:figure2}, a to e, insets). To ensure signal-to-noise optimization, the fundamental THz drive was slightly modified, as reflected in the spectral profiles of the driving field (dashed lines, Fig.~\ref{fig:figure2}, a to e). As the field increases, the photoresponse evolves from a broad, weakly structured low-field spectrum into a pronounced sequence of discrete high-frequency peaks (solid lines).
%and maximize signal-to-noise ratio
 
In the lower-field regime (Fig.~\ref{fig:figure2}e)---where transport remains interband-dominated and the time-domain signal undergoes a phase reversal---the photocurrent spectrum is relatively smooth and centered near the fundamental drive frequency $\omega_{0}$. However, even at these ``low" field strengths, the measured THz photocurrent spectrum (cyan solid line) is already highly nonlinear, i.e., significantly broader than the fundamental drive spectrum (black dashed line). This pronounced spectral broadening (cyan arrow, Fig.~\ref{fig:figure2}e) provides an unambiguous signature of a high-order nonlinear transport regime over the perturbative photoconductivity responses \cite{Ju2017Science} since the system is already generating new frequency components via the dominance of coherent multi-photon interband polarization.  
As the THz driving field amplitude increases from Fig.~\ref{fig:figure2}d to Fig.~\ref{fig:figure2}a, however, the increasing dominance of intraband ballistic acceleration drives highly non-equilibrium current trajectories. This non-perturbative interaction forces a dramatic redistribution of spectral weight, causing discrete features to accumulate near the high odd-order harmonics of the fundamental center frequency ($\omega_0$) while simultaneously narrowing the bandwidth at the fundamental order (blue arrow, Fig.~\ref{fig:figure2}a).

To inspect these THz high-harmonic-like photocurrent spectra with maximum visibility, Fig.~\ref{fig:figure2}f presents a log-scale comparison between the low-field crossover state ($E/E_\mathrm{max} = 0.39$, cyan) and the extreme strong-field limit ($E/E_\mathrm{max} = 1$, blue). Red dashed lines mark the odd-harmonic multipliers of the fundamental driving frequency. While the broadened low-field spectrum decays rapidly above $\sim 3~\mathrm{THz}$, the intense driving configuration reveals a striking, high-contrast comb of peaks resolving at least $9~\omega_0$ (($9\omega_0 \approx 5.2~\mathrm{THz}$)) and remains trackable up to the $13\mathrm{th}$ order ($\approx 7.5~\mathrm{THz}$) before meeting the experimental noise floor. Manifesting such an extended, broadband harmonic sequence requires the system to access extreme electronic correlations governed by high-order nonlinear conductivities ($\sigma^{(n)}$) ($n=3,5,7,9, 11\dots$)~\cite{Huang2026}.

Crucially, because our experimental observable is the rectified steady-state photocurrent extracted via sub-cycle interferometry rather than a directly radiated optical field, these features are most rigorously interpreted as multi-quantum coherences along the coherent control axis by varying the interpulse delay coordinate ($\tau$)~\cite{Huang2026}. Within a two-dimensional (2D) THz spectroscopy framework utilizing $(t, \tau)$ analysis~\cite{Mootz2022CommunPhys,Huang2026}, these high-order pathways map directly into the macroscopic DC transport channel at $\omega_{t} \approx 0~\mathrm{THz}$ (Supplementary). Their finite linewidths and approximate harmonic alignments reflect the mathematical projection of non-perturbative, multi-order electronic trajectories onto the $\omega_{\tau}$-axis, providing an electrical read-out of high-harmonic quantum dynamics embedded within a van der Waals nanodevice rather than perfectly discrete harmonic optical emission. 

%Note that their finite linewidth due to broadband THz driving and approximate harmonic positions reflect the projection of multi-order nonlinear photocurrent processes onto the inter-pulse delay coordinate, rather than perfectly discrete emitted harmonics.

\section{Gate tunability of high-order THz-field-driven photocurrents} 

We can classify the non-perturbative strong-field transport using two complementary control axes. First, at fixed bandgap, changing the THz electric-field amplitude tunes the Keldysh parameter $\gamma=\Delta/(2\hbar\Omega_{\mathrm R})$, as we have demonstrated in Fig.~\ref{fig:figure2}. Second, at fixed electric field, changing the electrostatic bandgap tunes the multiphoton parameter $M=\Delta/(\hbar\omega_0)$ while leaving the gap-independent, resonant adiabaticity parameter $\gamma/M=\omega_0/(2\Omega_{\mathrm R})$ unchanged. In the weak field limit or $\gamma \gg 1$, $N$-photon resonances occur when $M=N$, corresponding to $\Delta=N\hbar\omega_0$. 
Upon entering the non-perturbative regime ($\gamma < 1$), these simple integer resonance conditions are fundamentally altered by strong-field dynamics. Governed by the Rabi frequency ($\Omega_{\mathrm{R}}$) rather than bandgap ($\Delta$), these pristine integer conditions break down, shifting the resonance control from the integer value of $M$ alone to the field-dependent ratio $\gamma/M$. Combining both the Keldysh and multi-photon axes, K-axis and M-axis, therefore provides an experimental toolbox for navigating the entire strong-field non-perturbative phase space on demand (Supplementary Information, Section 4).

\begin{figure}[!ht] 
    \centering
    \includegraphics[width=1\linewidth]{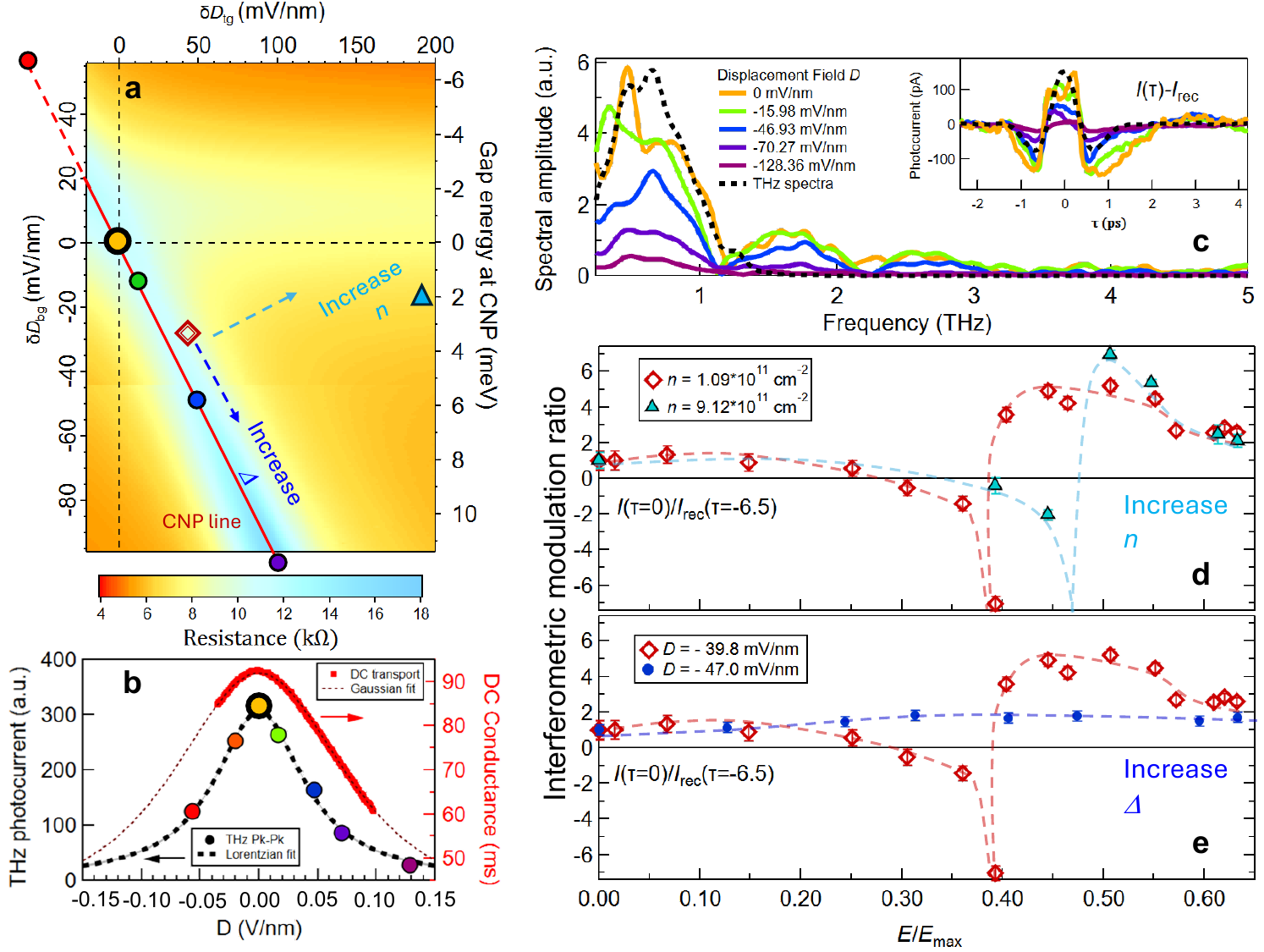}
    \caption{\textbf{Dual-gate electrostatic control and versatile tuning of non-perturbative THz transport.} 
    \textbf{a}, Two-dimensional resistance map of the BLG channel as a continuous function of top-gate ($\delta D_{\mathrm{tg}}$) and back-gate ($\delta D_{\mathrm{bg}}$) displacement fields. The solid red line marks the charge neutrality point (CNP) ridge, where the induced structural bandgap $\Delta$ scales up to $\sim 10~\mathrm{meV}$ (right axis). Colored circles and geometric markers define the precise operational coordinates evaluated across the orthogonal tuning trajectories.
    \textbf{b}, Peak-to-peak terahertz photocurrent (left axis, black circles) compared directly against the static DC transport conductance (right axis, red squares) along the CNP line as a function of the displacement field $D$. The localized terahertz photoresponse follows a sharp Lorentzian profile (black dashed line), while the macroscopic DC conductance exhibits a broader Gaussian distribution (red dashed line).
    \textbf{c}, Terahertz photocurrent spectral amplitudes obtained via Fourier transformation of the time-domain interferograms (inset) under maximum field driving for various displacement fields $D$ along the CNP line. The high-harmonic-like profile remains robustly preserved even as the bandgap opens far past the single-photon absorption threshold.
    \textbf{d}, Interferometric modulation ratio, $I_{\tau=0}/I_{\mathrm{rec}}$, versus the normalized driving field $E/E_{\mathrm{max}}$ for two distinct carrier densities: a lightly doped state ($n = 1.09 \times 10^{11}~\mathrm{cm}^{-2}$, red open diamonds) and an intensely doped state ($n = 9.12 \times 10^{11}~\mathrm{cm}^{-2}$, blue solid triangles). Increasing the carrier concentration shifts the non-perturbative zero-crossing to higher field thresholds via state-selective Pauli blocking.
    \textbf{e}, Comparison of the interferometric modulation ratio under orthogonal bandgap tuning at a constant background doping level. A moderate displacement field ($D = -39.8~\mathrm{mV\,nm^{-1}}$, red open diamonds) yields a high-contrast phase reversal, whereas an increased displacement field ($D = -47.0~\mathrm{mV\,nm^{-1}}$, blue solid circles) completely quenches the zero-crossing, pinning the response near the perturbative baseline.}
    \label{fig:figure3}
\end{figure}

Tuning the multiphoton parameter $M$ requires tuning the bandgap $\Delta$ while maintaining/controlling the Fermi energy/carrier density. By tuning the external top and back gate voltage across the dual gated BLG device (Fig.~\ref{fig:figure1}a), a wide phase space of parameters can be accessed~\cite{Zhang2009Nature,Taychatanapat2010PRL}. We can express the total carrier density ($n$, representing the doping gate action) and the net perpendicular displacement field ($D$, representing the inversion-symmetry-breaking gate action) through linear combinations: $n_e = \frac{1}{e}(\delta D_{\mathrm{tg}} + \delta D_{\mathrm{bg}})$ and $D = \frac{1}{2}(\delta D_{\mathrm{tg}} - \delta D_{\mathrm{bg}})$ (Supplementary Information). This 2D coordinate transformation (Fig.~\ref{fig:figure3}a) establishes an orthogonal control matrix defined by two primary lines: the prominent high-resistance charge neutrality point (CNP) ridge (red line), along which we continuously modulate the displacement field ($D$), and the perpendicular doping axis, which introduces a net charge carrier imbalance. This elevated resistance along the CNP line directly reflects the positioning of the Fermi energy within the tunable, field-induced bandgap, labeled in right axis, Fig.~\ref{fig:figure3}a, which minimizes the free carrier concentration. 
Along this CNP trajectory (red line, Fig.~\ref{fig:figure3}a), Fig.~\ref{fig:figure3}b maps the THz photoconductive response (solid circles) at the peak driving field across various displacement fields. This response is distributed symmetrically on both sides of the zero-gap point (yellow circles in Figs.~\ref{fig:figure3}b and \ref{fig:figure3}b, $D = 0~\mathrm{V\,nm^{-1}}$), allowing a direct benchmark against the baseline DC transport conductivity (red squares, Fig.~\ref{fig:figure3}b). 
%Both the THz peak-to-peak photocurrent (solid circles) and the DC transport conductance (thick red line) reach their absolute maxima at the gapless charge neutrality point (yellow circles, $D = 0~\mathrm{V\,nm^{-1}}$) and suppress systematically as a function of the widening electrostatic bandgap. 
Interestingly, the localized THz photoresponse is captured by a sharp Lorentzian profile (black dashed line) centered at this inversion-symmetric point, whereas the macroscopic DC conductance exhibits a broader Gaussian distribution (red dashed line). 
%Fig.~3A shows a 2D map of the DC transport measurement for our device. 
%The respective locations are marked on the 2D map in Figure 3(a). 

Figure~\ref{fig:figure3}c demonstrates the on-demand tunability of non-perturbative lightwave transport along the multiphoton control axis. Here, by pinning the driving THz field at its maximum amplitude, the gap-independent parameter $\gamma/M$ is held strictly constant while the displacement field ($D$) is systematically varied to adjust the bandgap ($\Delta$) with zero doping. The color-coded photocurrent spectra are obtained via Fourier transformation of the time-domain interferograms (Fig.~\ref{fig:figure3}c, inset), directly matching the coordinate markers along the CNP line in Fig.~\ref{fig:figure3}a. 
Interestingly, robust non-perturbative photoresponse persists even across exceptionally large energy barriers. Remarkably, clear spectral signatures remain trackable up to the highest displacement field of $D = -128.36~\mathrm{mV\,nm^{-1}}$, corresponding to an induced bandgap of $\Delta \approx 15~\mathrm{meV}$ (purple trace). This electrostatic gap is six times the fundamental driving photon energy ($\hbar\omega_0 \approx 2.44~\mathrm{meV}$ or $0.6~\mathrm{THz}$), demonstrating that intense lightwave driving enables non-perturbative carrier generation across an energy gap corresponding to a high-order multiphoton threshold.

We highlight one of the most distinguishing features of this multi-photon control axis: the characteristic lineshape of the high-order spectrum remains fundamentally invariant across the entire bandgap sweep (i.e., $M=0,1,\dots,6$). Specifically, although the overall spectral amplitude attenuates due to the expected reduction in carrier injection at larger gap sizes along the $M$-axis, the high-harmonic-like spectral architecture is preserved, distinct from the large spectral changes observed along the Keldysh axis. Comparing the large-gap profiles above $\omega_0$ (blue and purple traces, $\Delta \sim 6$ to $15~\mathrm{meV}$) to the small-gap configurations below $\omega_0$ (yellow and green traces, smaller than $2$~meV) reveals a uniform retention with similar discrete harmonic peak structure of the high-frequency spectral tail extending past $2-3~\mathrm{THz}$. This behavior reveals the hallmark of the deep non-perturbative regime where the integer condition $M=\Delta/(\hbar\omega_0)$ breaks down in the multi-photon resonant control axis and the invariant THz spectral shape is determined by $\frac{\gamma}{M}$ independent of $\Delta$. This behavior also stands in sharp contrast to the field-driven Keldysh tuning presented in Fig.~\ref{fig:figure2}, which mediated a distinct spectral transition from a broad, featureless low-field continuum into high-contrast comb peaks. Therefore, the intense field securely locks the system deep within the non-perturbative regime, while the orthogonal gate modulation along the $M$-axis provides an independent dial to sweep the same multi-photon resonant line at $M=6$ that regulates the amplitude of the high-order THz photocurrent. 
Moreover, the invariance of the photocurrent spectral profile as the bandgap is tuned above and below the fundamental frequency $\omega_0$ stands as another definitive hallmark of non-perturbative transport, distinct from the conventional bandgap controlled photocurrent conductivity spectra in the perturbative regime~\cite{Ju2017Science}.

Next we demonstrate deterministic control over the phase flips and zero-crossings of the high-order photocurrent by leveraging the gate-tuned competition between interband and intraband transport channels. First, transitioning perpendicular to the CNP line in Fig.~\ref{fig:figure3}a shifts the Fermi level via charge doping, moving from the lightly doped regime (red open diamond) to a heavily doped state (blue triangle). Figure~\ref{fig:figure3}d plots the interferometric modulation ratio, $I(\tau=0)/I_{\text{rec}}$, for these two doping levels, which span nearly an order of magnitude in carrier density ($n = 1.09 \times 10^{11}~\mathrm{cm}^{-2}$ versus $n = 9.12 \times 10^{11}~\mathrm{cm}^{-2}$) with only a slightly different bandgap setting. 
We observe a pronounced, high-field shift in the zero-crossing threshold with increased electron doping, where the onset of the phase flip migrates from $E/E_{\text{max}} \approx 0.25$ up to $\approx 0.48$. This trend is a direct consequence of Pauli blocking; as injected electrons fill the lower states of the conduction band, the phase space available for low-energy transitions is restricted. This effectively elevates the activation energy required for coherent sub-cycle steering, compressing the interband--intraband competition window and demanding significantly higher driving fields to initiate the non-perturbative phase reversal. 
Second, we explore the orthogonal control axis by tuning the displacement field ($D$) to dominantly alter the electronic bandgap ($\Delta$) while remaining close to the charge neutrality line (red diamonds and blue circles in Fig.~\ref{fig:figure3}a). At a moderate displacement field of $D = -39.8~\mathrm{mV\,nm^{-1}}$ ($\Delta \approx 3.5~\mathrm{meV}$, red diamonds in Fig.~\ref{fig:figure3}e), the system displays a sharp, high-contrast phase flip. Remarkably, increasing the displacement field to $D = -47.0~\mathrm{mV\,nm^{-1}}$ ($\Delta \approx 6.0~\mathrm{meV}$, blue circles in Fig.~\ref{fig:figure3}e) completely quenches this zero-crossing behavior across the entire field range used, flattening the modulation ratio $\approx 2$. This stark suppression highlights how opening an electrostatic bandgap penalizes the non-perturbative current generation, providing a gate switch to enable or disable lightwave current acceleration and phase flips. 

\section{Quantum-kinetic modeling of the multi-photon to lightwave acceleration crossover}

To identify the microscopic origin of the gate-tunable nonlinear THz photocurrent, we model the driven BLG using a quantum-kinetic two-band framework based on the semiconductor Bloch equations~\cite{McGouranDignam2017,Wilhelm2021SBE}. In this description, the total current is decomposed into interband and intraband contributions, $J_{\mathrm{tot}}(t,\tau)=J_{\mathrm{inter}}(t,\tau)+J_{\mathrm{intra}}(t,\tau)$, where the interband term arises from field-induced valence--conduction coherences and multiphoton excitation, while the intraband term reflects ballistic carrier acceleration within the bands. The detailed Hamiltonian, equations of motion, and current decomposition are given in the Supplementary Information (Section 5).

\begin{figure}[!ht] 
    \centering
    \includegraphics[width=0.8\linewidth]{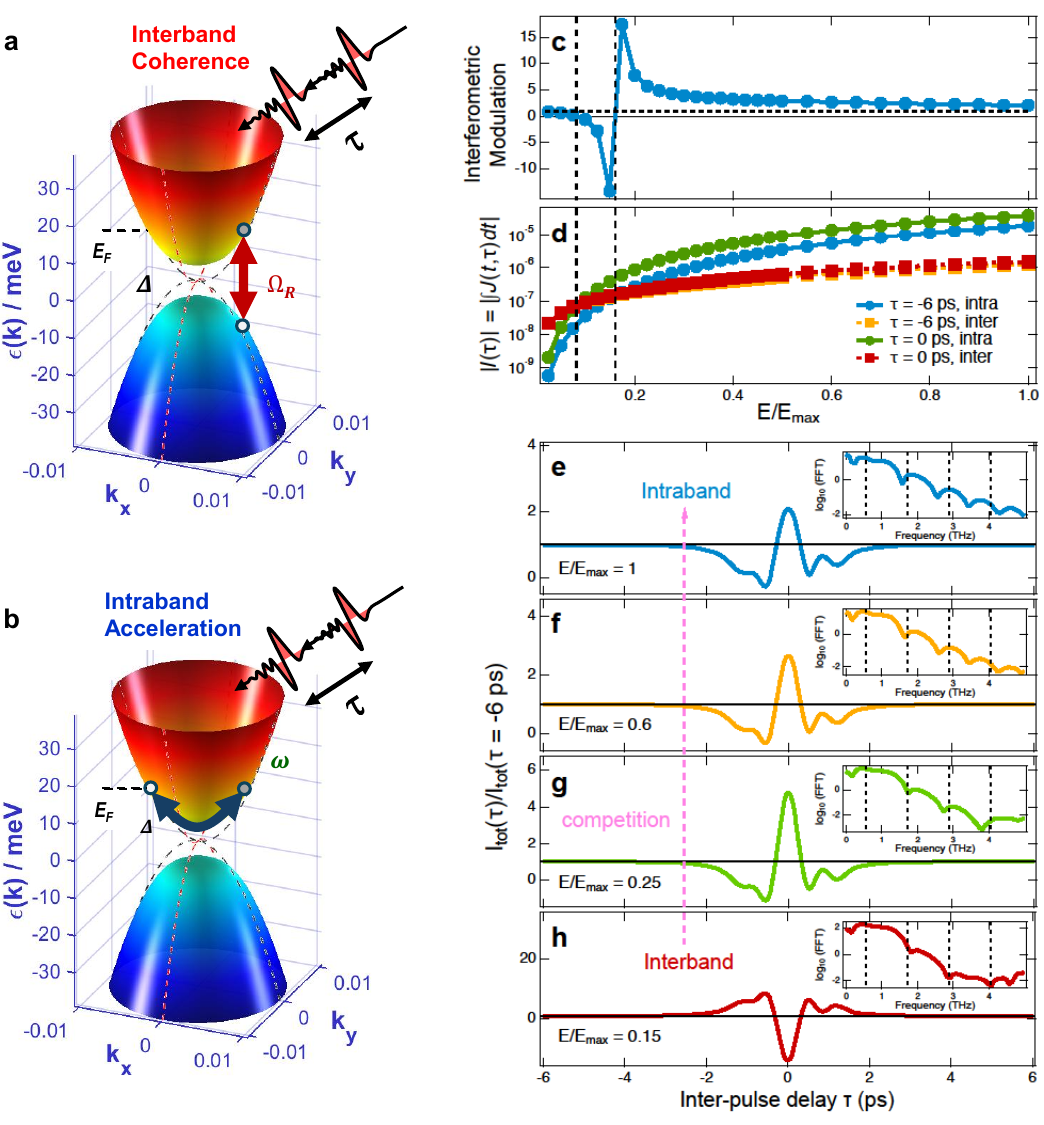}
    \caption{\textbf{Theory of the field-driven crossover between interband and intraband photocurrent channels in bilayer graphene.} \textbf{a,b}, Schematic illustration of the dominant microscopic processes induced by THz pulse pairs: interband multiphoton excitation and interband polarization (\textbf{a}) versus intraband lightwave acceleration (\textbf{b}). 
    \textbf{c}, Simulated field dependence of the photocurrent ratio, capturing the non-monotonic behavior and sign changes observed experimentally in Fig.~\ref{fig:figure1}e.
    \textbf{d}, Field dependence of the rectified interband and intraband current contributions for $\tau=-6$~ps and $\tau=0$, showing that the zero-crossings occur where the two channels become comparable and that the high-field regime is intraband-dominated.
    \textbf{e--h}, Simulated photocurrent ratio as a function of interpulse delay $\tau$ for representative driving fields selected from the distinct regimes in~\textbf{(c)}; corresponding delay-domain spectra are shown in the insets with vertical dashed lines indicating odd-order harmonics. Panels~\textbf{e--g} lie in the positive-ratio regimes, with \textbf{e} corresponding to the strongest driving field, whereas \textbf{h} represents the lowest-field negative-ratio regime. The spectra show the emergence of pronounced odd-order high-harmonic-generation-like spectral structure with increasing field strength, consistent with the experimental behavior in Fig.~\ref{fig:figure2}.}
    \label{fig4}
\end{figure}

Figures~\ref{fig4}a and \ref{fig4}b delineate the two overarching microscopic mechanisms governing this behavior. At lower driving fields, the response is governed primarily by interband processes because the finite gap and modest doping favor nonlinear interband polarization and multiphoton absorption over large-amplitude carrier acceleration. In this regime, the sign of the rectified current is governed primarily by the interband channel. As the THz field strength increases, however, the momentum excursion driven by the lightwave field grows rapidly and the intraband contribution rises more strongly. The measured rectification therefore reflects a competition between these two channels rather than a single nonlinear mechanism. 

Figures~\ref{fig4}c and \ref{fig4}d establish the microscopic origin of the nonlinear field dependence observed experimentally in Figs.~\ref{fig:figure1}d and \ref{fig:figure1}e. Figure~\ref{fig4}c captures the measured evolution of the photocurrent ratio with driving field, while Fig.~\ref{fig4}d decomposes the rectified signal into interband and intraband contributions for $\tau=-6$~ps and $\tau=0$. The calculations show that the INTENSE response is governed by a field-driven crossover between interband-dominated response and intraband-dominated transport. At low fields, the rectified photocurrent is controlled primarily by interband processes, but as the THz field increases intraband current grows more rapidly and eventually exceeds the interband contribution. This crossover explains both the phase reversal in Fig.~\ref{fig:figure1}d and the non-monotonic behavior in Fig.~\ref{fig:figure1}e. The two zero-crossings mark compensation points (vertical dashed lines) at which interband and intraband rectified currents become comparable in magnitude (Fig.~\ref{fig4}d), causing the net ratio signal to vanish before changing sign. In the intermediate regime, the response is therefore exceptionally sensitive to THz field strength, whereas at higher fields the intraband channel dominates the rectification current channel and restores an in-phase response with a qualitatively different interferogram profile. The comparison between overlapping and well-separated pulses further shows that pulse overlap enhances nonlinear mixing, while the separated-pulse limit more directly reveals the intrinsic balance between the two current channels.

Figures~\ref{fig4}e--\ref{fig4}h provide the theoretical counterpart to the field-dependent delay-domain photocurrent response measured in Fig.~\ref{fig:figure2}. The simulations show the photocurrent ratio as a function of interpulse delay $\tau$ for representative driving fields spanning the distinct regimes identified in Fig.~\ref{fig4}c, while the insets display the corresponding delay-domain spectra. Panels~\ref{fig4}e--\ref{fig4}G lie in the positive-ratio regimes, evolving from the strongly driven intraband-dominated response in Fig.~\ref{fig4}e toward the crossover regime at lower fields, whereas Fig.~\ref{fig4}h represents the lowest-field negative-ratio regime in which the rectified response has reversed sign. This progression shows that the spectral evolution is tied directly to the same field-driven competition between interband and intraband processes discussed above. At lower fields, where the response remains interband dominated and the ratio can become negative, the spectrum exhibits a broader fundamental peak and is otherwise less structured, consistent with Fig.~\ref{fig:figure2}e. As the field increases, strong intraband acceleration together with repeated interband polarization produces increasingly nonlinear current dynamics, causing spectral weight in the spectra to accumulate near odd multiples of the fundamental frequency. The simulated evaluation from Fig.~\ref{fig4}h to Fig.~\ref{fig4}e therefore mirrors the experimental trend in Fig.~\ref{fig:figure2} from a broader low-field response to a pronounced odd-order high-harmonic-generation-like spectral pattern at strong driving. Because the measured quantity is the rectified photocurrent extracted from the interferogram rather than the directly emitted AC harmonic field, these features are most accurately described as odd-order high-harmonic generation-like spectral signatures in the photocurrent channel. 
Instead of perfectly discrete harmonic profiles, the spectra give way to peaks with finite linewidths and approximate harmonic positions dependent on field strength that reflect the projection of higher-order nonlinear quantum coherence and pathways onto the delay coordinate~\cite{Huang2026}.

The above picture also clarifies the gate tuning observed and discussed in Fig.~\ref{fig:figure3}. Increasing carrier density suppresses interband excitation through Pauli blocking and shifts the crossover to higher field, while increasing the displacement-field-induced gap raises the effective threshold for multiphoton interband processes. Nevertheless, under sufficiently strong THz driving, the intraband and strongly nonlinear interband channels remain active even when the gap exceeds the fundamental photon energy by several times, allowing the odd-order HHG-like spectral structure to persist. Overall, the theory shows that the gate-tunable nonlinear photocurrent in BLG is controlled by a field-driven crossover from an interband-dominated multiphoton regime to an intraband-dominated lightwave-acceleration regime, with the phase flips and zero-crossings providing a direct experimental signature of where these two channels balance each other.

\section{Conclusion}
We have demonstrated the gate and field tuning of non-perturbative lightwave photocurrents in bilayer graphene via sub-cycle intense terahertz interferometry, uncovering a striking landscape of photocurrent spectra, field-dependent phase reversals and zero-crossings. Framed within a THz 2D spectroscopy framework, these signatures manifest along the photocurrent rectification axis as a high-contrast comb of multi-quantum coherences extending up to the exceptionally high order, rooted in a microscopic crossover from interband multiphoton transitions to intraband lightwave acceleration. 
The ability to control non-perturbative photocurrent transport regimes on-demand can be extended to other atomically thin device architectures such as twisted graphene bilayers, and opens immediate avenues for gate- and field-programmable 2D lightwave devices and quantum simulators, where electronic pathways can be steered on sub-cycle, petahertz timescales and atomic length scales. 
%By bridging attosecond strong-field physics with solid-state nanoelectronics, this platform opens immediate avenues for field-programmable petahertz devices and lightwave-driven quantum simulators.

%\begin{figure}[!ht] 
 %   \centering
  %  \includegraphics[width=1\linewidth]{Figures/Fig_S1_v3.png}
   % \caption{Enter Caption}
    %\label{fig:placeholder}
%\end{figure}

%\begin{figure}[!ht] 
 %   \centering
  %  \includegraphics[width=1\linewidth]{Figures/Fig_S2_v3.png}
   % \caption{Enter Caption}
    %\label{fig:placeholder}
%\end{figure}

%\begin{figure}[!ht] 
 %   \centering
  %  \includegraphics[width=1\linewidth]{Figures/Fig_S4_v3.png}
   % \caption{Enter Caption}
    %\label{fig:placeholder}
%\end{figure}

\bmhead{Acknowledgements}
The nonlinear THz photocurrent spectroscopy measurement was supported by the US Department of Energy, Office of Basic Energy Science, Division of Materials Sciences and Engineering (Ames National Laboratory is operated for the US Department of Energy by Iowa State University under contract no. DE-AC02-07CH11358). BLG device fabrication and basic characterization at MIT were supported by NSF grant DMR-2414725.

\bibliography{ref}% common bib file

\end{document}